\documentclass[a4paper]{article}%
\usepackage{amsmath}
\usepackage{amssymb}
\usepackage{amsfonts}
\usepackage{graphicx}%
\begin{document}

\title{On the Regularisation in J-matrix Methods}
\author{J. Broeckhove$^{1}$, V.S. Vasilevsky$^{2}$, F. Arickx$^{1}$, and A.M.
Sytcheva$^{1}$\\$^{1}$University of Antwerp, \\Group Computational Modeling and Programming, \\Antwerp, Belgium\\$^{2}$Bogolyubov Institute for Theoretical Physics,\\Kiev, Ukraine}
\date{}
\maketitle

\begin{abstract}
We investigate the effects of the regularization procedure used in the
J-Matrix method. We show that it influences the convergence, and propose an
alternative regularization approach.We explicitly perform some model
calculations to demonstrate the improvement.

\end{abstract}

\section{Introduction}

The J-Matrix (JM) method has become a popular tool in solving quantum
mechanical scattering problems in atomic, nuclear and molecular systems. It is
based on a square-integrable form of the wave function. The basis used for the
representation of the wave function has to reduce the reference Hamiltonian,
which is responsible for the asymptotic behavior of the solutions, to a
tridiagonal or Jacobi matrix form. The scattering boundary conditions can then
be expressed in terms of the expansion coefficients of the wave function. The
Schr\"{o}dinger equation has to be solved by matching the interior to the
asymptotic region, which can now be done on the expansion coefficients. This
leads to a set of linear equations, the solutions of which are the expansion
coefficients of the interior representation, and the scattering parameters
(phase shift or $S$-matrix).

The JM approach proved to be succesful for 2-, 3- and many-particle systems in
atomic and molecular
\cite{kn:Heller2,Reinhardt72,1995JPhB...28L.139K,1976PhRvA..14.2159B,1993PAN....56..886K}%
, and nuclear physics
\cite{2001JETP...92..789K,vasil_rybk89,mirror_react90,kn:Vasil92,kn:vasil97e,kn:ITP+RUCA1,kn:ITP+RUCA2,kn:ITP+RUCA3}%

Despite the successes, a slow convergence as a function of the number of basis
functions in the internal region leads to the need of large matrices. The
calculation of matrix elements of the potential energy operator usually
represents the main computational cost in a realistic many-body application.
An improvement on the method leading to faster convergence with a smaller size
of the Hamiltonian matrix is therefore necessary. There have already been
attempts, based on asymptotic properties of the potential
\cite{2002PhRvL..88a0404V}, that led to sometimes dramatic reductions of the
size of the modelspace. Other possibilities lie in computational methods for
faster evaluation of potential matrix elements. Still a number of problems
remain to solve the convergence issue.

In this paper we analyze the behavior of the irregular (or Neumann-like)
solution of the free-motion Schr\"{o}dinger equation, and the corresponding
expansion coefficients. Because of the square-integrable representation, a
regularizing boundary condition has to be introduced to solve this equation.
The standard regularization considered in the JM approach
\cite{kn:Heller1,kn:SmirnovE,kn:Yamani} will be shown to put a limit on the
minimal size of the modelspace, and thus to have an important impact on the
convergence. An analysis of the regularized asymptotic solution leads to a new
regularization procedure that improves the size limitation. The enhancement in
convergence of this procedure will be explicitly demonstrated on a number of
potential problems.

We consider a JM formulation based on the oscillator basis, and a free-motion
Hamiltonian. The analysis and the new regularization procedure can be easily
extended to other bases, with analogous conclusions.

\section{J-Matrix Methods}

A Schr\"{o}dinger equation for a continuum wave function with a spherically
symmetric, non-coulombic potential,%
\begin{equation}
-\frac{\hbar^{2}}{2m}\left(  \frac{1}{r^{2}}\frac{\partial}{\partial
r}\!\!\left(  r^{2}\frac{\partial}{\partial r}\right)  \!\!-\!\!\frac
{l(l+1)}{\hbar^{2}r^{2}}\!\!+\!\!V(r)\!\!-\!\!E\!\right)  \!\!\Psi
_{l}(r)\!\!=\!\!0 \label{eq:schroedinger}%
\end{equation}
must have a solution that is matched asymptotically with the free-space Bessel
and Neumann functions%
\begin{equation}
\Psi_{l}(r\rightarrow\infty)\rightarrow\sqrt{\frac{2}{\pi}}\,j_{l}%
(kr)-\tan\delta_{l}(k)\,\ \sqrt{\frac{2}{\pi}}\,n_{l}(kr)
\label{eq:coordinateasymptotic}%
\end{equation}
where $k^{2}=\frac{2mE}{\hslash^{2}}.$This match of a solution in the
interaction region, where the effect of the potential is felt, with the
asymptotic reference states determines the phase shift at momentum $k$
corresponding to energy $E$. We use the traditional spherical Bessel and
spherical Neumann function definitions \cite{kn:Newton,kn:MorseFeshbach} with
the delta-function normalization.

The JM method reformulates this problem into an algebraic setting by the use
of a basis functions. In our case these are the harmonic oscillator
functions\thinspace\thinspace%
\begin{align}
\Phi_{nl}(r\,|\,b)  &  =(-1)^{n}\frac{1}{b^{3/2}}N_{nl}\,(\frac{r}{b}%
)^{l}L_{n}^{l+1/2}\left(  (\frac{r}{b})^{2}\right)  \exp\left(  -\frac{1}%
{2}(\frac{r}{b})^{2}\right) \\
N_{nl}  &  =\sqrt{\frac{2\Gamma\left(  n+1\right)  }{\Gamma\left(
n+l+3/2\right)  }}\nonumber
\end{align}
where the $L_{n}^{\alpha}(r)$ are Laguerre polynomials, $N_{nl}$ is a
norm-factor and $b$ is the oscillator length. We take a functional notation in
which the variable of the function is separated from the parameter by a
\textquotedblleft$|$\textquotedblright; it will be used througout the paper.
While it is straightforward to turn equation (\ref{eq:schroedinger}), using
the superposition ansatz
\begin{equation}
\Psi_{l}(r|k)=\sum_{n=0}^{\infty}C_{nl}(k,b)\Phi_{nl}(r\,|\,b)
\label{eq:OscExp}%
\end{equation}
into a matrix equation, more care must be taken when mapping the boundary
condition (\ref{eq:coordinateasymptotic}). The Bessel and Neumann functions
are the solutions of the free-space reference Hamiltonian.

The regular Bessel solution%

\begin{align}
\left(  T_{l}-E\right)  \,\!\!\Psi_{l}^{B}(r|k)\!\!  &  =\!\!0
\label{eq:bessel}\\
\Psi_{l}^{B}(r\,|\,k)\!\!  &  =\sqrt{\frac{2}{\pi}}\,j_{l}(kr)
\end{align}
presents no problems and can be reconstructed as the solution of%

\begin{equation}
\sum_{m=0}^{\infty}\left\langle \Phi_{nl}(b)|T_{l}-E|\Phi_{ml}(b)\right\rangle
C_{ml}^{B}(k,b)=0
\end{equation}
where the parameter of the basis states in brackets is indicated as a variable
notation; this notation is used throughout the paper. This is a three-term
recurrence relation that can be solved explicitly
\cite{kn:Heller1,kn:SmirnovE,kn:Yamani}:%
\begin{align}
C_{nl}^{B}\left(  k,b\right)   &  =N_{nl}\,b^{3/2}\,(kb)^{l}\exp\left(
-\frac{1}{2}(kb)^{2}\right)  \,L_{n}^{l+1/2}\left(  (kb)^{2}\right)
\nonumber\\
&  =\frac{2}{N_{nl}}\frac{b^{3/2}}{\Gamma(l+\frac{3}{2})}(kb)^{l}\exp
(-\frac{(kb)^{2}}{2})_{1}F_{1}\left(  -n,\,l+\frac{3}{2}\,;(kb)^{2}\right)
\label{eq:C021}%
\end{align}
whre the the $_{1}F_{1}$ stands for Kummer's function \cite{kn:abra}. For
large $n$, the asymptotic behavior of the\ coefficients is given by%
\begin{equation}
\ C_{nl}^{B}(k,b)\rightarrow b\sqrt{2R_{nl}}\Psi_{l}^{B}(R_{nl\,}|\,k)
\label{eq:AsymCoefBess}%
\end{equation}
where the $R_{nl\,}=b\sqrt{4n+2l+3}$ are the oscillator turning points.

The Neumann solution is irregular%

\begin{align}
\left(  T_{l}-E\right)  \,\!\!\Psi_{l}^{N}(r|k)\!\!  &  =\!\!0\\
\Psi_{l}^{N}(r\,|\,k)\!\!  &  =\sqrt{\frac{2}{\pi}}\,n_{l}(kr)
\end{align}
and cannot be mapped directly because the oscillator states cannot represent
the singular behavior at the origin. This is not a fundamental problem because
we only require these reference solutions to formulate the asymptotic
conditions i.e. for $r\rightarrow\infty$. An equivalent solution, i.e. one
that behaves asymptotically as the Neumann function but is regular at the
origin, will do just as well. In some applications one uses a simple cutoff
procedure. In the JM method it is of course important that the procedure
translates easily to the algebraic representation of the problem. One takes
the solution of%

\begin{align}
\left(  T_{l}-E\right)  \overline{\Psi}_{l}^{N}\left(  r\,|\,k,b\right)   &
=\beta\,\Phi_{0l}\left(  r\,|\,b\right)  \label{eq:regNeumm}\\
\overline{\Psi}_{l}^{N}\left(  r\,|\,k,b\right)   &  \rightarrow\sqrt{\frac
{2}{\pi}}\,n_{l}(kr)\,\,\,\,\,\text{for}\,\,\,r\rightarrow\infty
\end{align}
The regularized Neumann function can be determined from this equation using
the Green's function technique \cite{kn:Heller1},\cite{kn:Yamani} (see also
next section). The coefficient $\beta$\ is fixed by the boundary condition.
This regularized Neumann function can be reconstructed as a solution of the equation%

\begin{equation}
\sum_{m=0}^{\infty}\left\langle \Phi_{nl}(b)|T_{l}-E|\Phi_{ml}(b)\right\rangle
C_{ml}^{N}(k,b)=\beta\ \delta_{n,0} \label{eq:C020}%
\end{equation}
It is a three-term recurrence relation with a simple inhomogeneous
contribution and can again be solved explicitly%

\begin{equation}
C_{nl}^{N}\left(  k,b\right)  =\frac{(-)^{l+1}N_{nl}b^{\frac{3}{2}}}%
{\Gamma\left(  -l+\frac{1}{2}\right)  }(kb)^{-l-1}\exp\left(  -\frac{(kb)^{2}%
}{2}\right)  \,_{1}F_{1}\left(  -n-l-\frac{1}{2},-l+\frac{1}{2};(kb)^{2}%
\right)  \label{eq:C022}%
\end{equation}
For large n, the asymptotic behavior is%
\begin{equation}
\ C_{nl}^{N}(k,b)\rightarrow b\sqrt{2R_{nl}}\overline{\Psi}_{l}^{N}\left(
r\,|\,k,b\right)  \rightarrow b\sqrt{2R_{nl}}\Psi_{l}^{N}(R_{nl\,}|\,k)
\label{eq:AsymCoeffNeum}%
\end{equation}
We use these asymptotic values (\ref{eq:AsymCoefBess}) and
(\ref{eq:AsymCoeffNeum}) to compute the $C_{nl}^{B}(k,b)$\ and $C_{nl}%
^{N}(k,b)$\ coefficients by seeding the recurrence relations at some large
index value, and recurring back towards small $n$. This turns out to be faster
and numerically more efficient than directly evaluating the analytical expressions.

Thus we are led to a formulation of the scattering problem where we represent
the solution as follows. The wave function has two essential components. In
the internal region the effects of the potential are felt. We assume it has a
finite range and that this component can be approximated by a finite
combination of the $L^{2}$ basis states. In the asymptotic region the solution
must match the boundary condition which we now express with the Bessel and
regularized Neumann functions.%

\begin{align}
\Psi_{l}\left(  r|k\right)   &  =\Psi_{l}^{I}(r|k)\,+\Psi_{l}^{B}%
(r|k)-\tan\delta_{l}\overline{\,\Psi}_{l}^{N}\left(  r|k,b\right)
\,\label{eq:C000}\\
\Psi_{l}^{I}(r|k)  &  \rightarrow0\,\,\,\,\,\,\text{for}\,\,\,r\rightarrow\infty
\end{align}
In the algebraic representation this becomes%

\begin{align}
C_{nl}(k,b) &  =C_{nl}^{I}(k,b)+C_{nl}^{B}(k,b)-\tan\delta_{l}\,C_{nl}%
^{N}(k,b)\\
C_{nl}^{I}(k,b) &  \rightarrow0\,\,\,\,\,\,\text{for}\,\,\,n\rightarrow\infty
\end{align}
The equation for the internal wavefunction and the phase shift is obtained by
inserting (\ref{eq:C000}) into the Schr\"{o}dinger equation, projected onto
the basis set:%

\begin{equation}
\left\langle \Phi_{nl}(b)|T_{l}+V-E|\Psi_{l}^{I}(k)+\Psi_{l}^{B}(k)-\tan
\delta_{l}\,\overline{\Psi}_{l}^{N}(k,b)\right\rangle =0\,\;\;\text{for\,all}\,n
\end{equation}
Taking advantage of the properties of the asymptotic reference states
(\ref{eq:bessel}) and (\ref{eq:regNeumm}) this simplifies to:%

\begin{align}
&  \left\langle \Phi_{nl}(b)|T_{l}+V-E|\Psi_{l}^{I}(k)\right\rangle
-\tan\delta_{l}\left(  \beta\delta_{n0}+\left\langle \Phi_{nl}(b)|V|\overline
{\Psi}_{l}^{N}(k,b)\right\rangle \right)  \nonumber\\
&  =-\left\langle \Phi_{nl}(b)|V|\Psi_{l}^{B}(k)\right\rangle
\end{align}
In the approach of \cite{kn:Heller1,kn:Yamani} the algebraic Schr\"{o}dinger
equation is approximated by assuming that the interaction matrix for potential
V can be truncated at some large $n=N$. This $N$ defines a sharp boundary
between the internal and asymptotic regions in the coefficient space. The
resulting problem can be solved with $C_{nl}^{I}(k,b)=0$\ for $n\,\geq N$ .
One has an $N+1$ by $N+1$\ matrix equation in the unknowns $\{C_{0l}%
^{I},C_{1,l}^{I},\ldots,C_{N-1l}^{I},\tan\delta_{l}\}$:%

\begin{align}
&  \sum_{m=0}^{N-1}\left\langle \Phi_{nl}(b)|T_{l}+V-E|\Phi_{ml}%
(b)\right\rangle C_{ml}^{I}(k,b)\nonumber\\
&  \;\;\;\;\;\;-\tan\delta_{l}(k)\left(  \beta\delta_{n0}+\sum_{m=0}%
^{N-1}\left\langle \Phi_{nl}(b)|V|\Phi_{ml}(b)\right\rangle C_{ml}%
^{N}(k,b)\right) \nonumber\\
&  =-\sum_{m=0}^{N-1}\left\langle \Phi_{nl}(b)|V|\Phi_{ml}(b)\right\rangle
C_{ml}^{B}(k,b)
\end{align}
Convergence of the results is achieved by extending the interaction region
i.e. increasing $N$.

\section{The Regularized Neumann function}

The regularized Neumann function is a solution of the inhomogeneous
differential equation (see details for the definition of $\overline{\Psi}%
_{l}^{N}\left(  r|k,b\right)  $ in \cite{kn:Heller1,kn:SmirnovE,kn:Yamani}):%
\begin{equation}
\left(  T_{l}-E\right)  \overline{\Psi}_{l}^{N}\left(  r\,|\,k,b\right)
=\beta\,\Phi_{0l}\left(  r\,|\,b\right)  \label{eq:C101}%
\end{equation}
where%
\begin{equation}
\beta=-\frac{\hbar^{2}}{m\pi k}\frac{1}{\Phi_{0l}\left(  k|1/b\right)  }.
\label{eq:C102}%
\end{equation}
The integral representation of $\overline{\Psi}_{L}^{N}\left(  r|k,b\right)  $
in coordinate space can be written as:%

\begin{equation}
\overline{\Psi}_{l}^{N}\left(  r\,|\,k,b\right)  =\beta%
{\displaystyle\int\limits_{0}^{\infty}}
G(r,r^{\prime})\,\Phi_{0l}(r^{\prime}|\,b)dr^{\prime}%
\end{equation}
where $G(r,r^{\prime})$ is the Green's function, explicitized in this case as:%
\begin{equation}
G(r,r^{\prime})=\frac{-\beta mk}{\hbar^{2}}j_{l}(kr_{<})\,n_{l}(kr_{>})r^{2}%
\end{equation}
In what follows we will restrict ourselves to the case of zero angular
momentum. The regularized Neumann function can then be calculated analytically
starting from the expression above, by straightforward integration of a
combination of sines, cosines and gaussians. One finds%

\begin{align}
\overline{\Psi}_{0}^{N}\left(  r\,|\,k,b\right)  =-\frac{1}{\sqrt{2\pi}kr}[
&  \exp(-ikr)\operatorname{erf}\left(  \frac{kr}{\sqrt{2}kb}-\frac{ikb}%
{\sqrt{2}}\right) \nonumber\\
+  &  \exp(ikr)\operatorname{erf}\left(  \frac{kr}{\sqrt{2}kb}+\frac
{ikb}{\sqrt{2}}\right)  ] \label{eq:C105}%
\end{align}
From this expression, we can see that the oscillator length has a strong
impact on the behavior of $\overline{\Psi}_{0}^{N}\left(  r\,|\,k,b\right)  $.
Indeed, when $kb$ tends to zero, and if we take only the leading term in the
series approximation of the complex error function for small complex argument,
then%
\begin{align}
\overline{\Psi}_{0}^{N}\left(  r\,|\,k,b\right)   &  \simeq-\frac{2}%
{\sqrt{2\pi}kr}\cos(kr)\cdot\operatorname{erf}\left(  \frac{kr}{\sqrt{2}%
kb}\right) \\
&  =\Psi_{0}^{N}\left(  r\,|\,k\right)  \cdot\operatorname{erf}\left(
\frac{r}{\sqrt{2}b}\right)  .\nonumber
\end{align}
Thus the larger is $kr/kb=r/b$, the closer the regularized function
$\overline{\Psi}_{0}^{N}(r|k,b)$ is to the original Neumann function $\Psi
_{0}^{N}\left(  r|k\right)  $. For instance, when $r\geq5b$, the functions
$\overline{\Psi}_{0}^{N}(r|k,b)$ and $\Psi_{0}^{N}\left(  r|k\right)  $ are
equal within single precision computation. When $kb$ becomes large then%
\begin{align}
\overline{\Psi}_{0}^{N}(r|k,b)  &  \approx-\sqrt{\frac{2}{\pi}}\frac{1}%
{kr}\cos(kr)+\frac{2}{\pi}\frac{kb}{(kr)^{2}+(kb)^{4}}\exp\left\{  -\frac
{1}{2}\frac{(kr)^{2}}{(kb)^{2}}-\frac{1}{2}(kb)^{2}\right\} \nonumber\\
&  =\Psi_{0}^{(N)}(r|k)+\frac{2}{\pi}\frac{kb}{(kr)^{2}+(kb)^{4}}\exp\left\{
-\frac{1}{2}\frac{(kr)^{2}}{(kb)^{2}}-\frac{1}{2}(kb)^{2}\right\}
\label{eq:C106}%
\end{align}
In this case $kr$ should be large compared to $kb$ to suppress the second term
in this expression, so that the function $\overline{\Psi}_{0}^{N}(r|k,b)$
coincides with $\Psi_{0}^{N}(r|k)$. The larger is $kb$, the larger the
coordinate $r$ has to be to reduce the difference between $\overline{\Psi}%
_{0}^{N}(r|k,b)$ and $\Psi_{0}^{N}(r|k)$. For values of $r\leq b$ , the
regularized Neumann function can be represented as%

\begin{equation}
\overline{\Psi}_{0}^{N}\left(  r\,|\,k,b\right)  =-\frac{1}{\pi kb}\left[
-2\exp\left(  \frac{(kb)^{2}}{2}\right)  -i\sqrt{2\pi}kb\operatorname{erf}%
\left(  \frac{ikb}{\sqrt{2}}\right)  \right]  +\ldots
\end{equation}
by taking the leading term of the Taylor expansion of (\ref{eq:C105}).\ Unlike
the appearance, this expression is real because the error function is
imaginary at an imaginary argument. One remarks (confirmed later by figure
\ref{Fig:PsiMinus1}) that in the limit for $kb\rightarrow0$ one has%

\begin{equation}
\overline{\Psi}_{0}^{N}\left(  0\,|\,k,b\right)  \simeq-\frac{1}%
{kb}\longrightarrow-\infty\,\,\,\,\,\,\,\,\,\,\text{for}\,\,\,kb\rightarrow0
\end{equation}
whereas when $kb\rightarrow\infty$, the function behaves as:%

\begin{equation}
\overline{\Psi}_{0}^{N}(0|k,b)\simeq\frac{\exp((kb)^{2}/2)}{kb}\longrightarrow
+\infty\,\,\,\,\,\,\,\,\,\text{\thinspace for}\,\,\,kb\rightarrow\infty
\end{equation}
The second situation in particular may be the source of numerical difficulties
in actual calculations. Perhaps more to the point for the JM method is the
fact that this behavior is reflected in the corresponding expansion
coefficients $C_{n0}^{N}(k,b)$ for $kb\ll1$ and for\ $kb\gg1$. Those
coefficients then become very large at small values of $n$. From the
analytical expressions (\ref{eq:C022}) for $C_{00}^{N}(k,b)$, and using the
expressions in \cite{kn:abra} for the limiting behavior of Kummer's function,
one finds for instance:
\begin{equation}
C_{00}^{N}(k,b)=O\left(  \frac{1}{kb}\right)  \,\longrightarrow+\infty
\,\,\,\,\,\,\,\text{for\thinspace}\,\,kb\rightarrow0\,
\end{equation}
and%

\begin{equation}
C_{00}^{N}(k,b)=O\left(  \frac{\exp\left(  \frac{1}{2}\left(  bk\right)
^{2}\right)  }{kb}\right)  \,\longrightarrow+\infty\,\,\,\,\,\,\,\text{for}%
\,\,\,kb\rightarrow\infty\,
\end{equation}
Care is needed in these situations to avoid numerical difficulties associated
with the large values of these coefficients..%

\begin{figure}
[ptb]
\begin{center}
\includegraphics[
trim=0.737460in 0.000000in 0.263379in 0.369327in,
natheight=5.947300in,
natwidth=6.753300in,
height=10.6976cm,
width=11.0314cm
]%
{PsiMinus.eps}%
\caption{Neumann $\Psi_{l}^{N}$ and regularized $\overline{\Psi}_{l}^{N}$
functions, for $l=0$, $k=1$ and various oscillator radii $b$. ($\overline
{\Psi}_{l}^{N}$ for $b=0.1$ fm coincides with $\Psi_{l}^{N}$ on this scale)}%
\label{Fig:PsiMinus1}%
\end{center}
\end{figure}

In figure \ref{Fig:PsiMinus1} we display the behavior\ of the $\overline{\Psi
}_{0}^{N}(r|k,b)$ function for $k=1$ and for values of the oscillator radii
$b=0.1$ up to $b=4.0$ fm. One clearly notices that for small values of $b$ the
functions $\overline{\Psi}_{0}^{N}(r|k,b)$ and $\Psi_{0}^{N}(r|k)$ are very
close to each other in the whole range of $r$, with the exception of small
$r$. The larger is $b$ (or better, in general, $kb$), the larger is the value
of the coordinate $r$ where the functions $\overline{\Psi}_{0}^{N}(r|k,b)$ and
$\Psi_{0}^{N}(r|k)$ are equal within required numerical precision.

The present analysis has shown that the combined parameters for the momentum
$k$, and the oscillator length $b$ (more specifically the product $kb$) have a
great impact on the behavior of the function $\overline{\Psi}_{l}^{N}(r|k,b)$.
For increasing $b$, given a fixed $k$, there is an increasing region of the
coordinate $r$ where the regularized and original Neumann functions differ
considerably. Small values of $b$ thus seem preferable for obtaining a faster
convergence in the JM method.

\section{The J-Matrix method revisited\label{sect:JMrevisited}}

In this section we will define new expansion coefficients for the regularized
Neumann functions in order to increase the convergence rate of the JM method.
For this aim we rewrite the equation (\ref{eq:C101}) in the following way%
\begin{equation}
\left(  T_{l}-E\right)  \overline{\Psi}_{l}^{N}\left(  r|k,b_{0}\right)
=\beta_{0}\,\Phi_{0l}\left(  r|b_{0}\right)  \label{eq:newregneum}%
\end{equation}
We have chosen an oscillator state $\Phi_{0l}\left(  r|b_{0}\right)  $ for the
regularization procedure that has a different oscillator radius $b_{0}$ than
the oscillator radius $b$ of the basis for the expansion (\ref{eq:OscExp}),
and a corresponding $\beta_{0}$. The expansion coefficients of the (new)
regularized Neumann function (\ref{eq:newregneum})%

\begin{equation}
C_{nl}^{N,b}\left(  k,b_{0}\right)  =\left\langle \Phi_{nl}(b)\,|\,\overline
{\Psi}_{l}^{N}\left(  k,b_{0}\right)  \right\rangle
\end{equation}
now satisfy the following set of algebraic inhomogeneous equations%

\begin{equation}
\sum_{m=0}^{\infty}\,\left\langle \Phi_{nl}(b)\left\vert T_{l}-E\,\right\vert
\Phi_{ml}(b)\right\rangle C_{ml}^{N,b}\left(  k,b_{0}\right)  =\beta
_{0}\left\langle \Phi_{nl}(b)\,|\Phi_{0l}(b_{0})\right\rangle
\label{eq:newregneumrecc}%
\end{equation}
i.e. a three-term inhomogeneous recurrence relation, with a source containing
the overlap $\left\langle \Phi_{nl}(b)\,|\Phi_{0l}(b_{0})\right\rangle $. The
latter can be easily calculated (see Appendix \ref{Sect:Appendix}):%
\begin{equation}
\left\langle \Phi_{nl}(b)\,|\,\Phi_{0l}(b_{0}\right\rangle =\left(
\frac{2bb_{0}}{b^{2}-b_{0}^{2}}\right)  ^{j}\sqrt{\frac{\Gamma\left(
n+j\right)  }{n!\Gamma\left(  j\right)  }}\,\left(  \frac{b^{2}-b_{0}^{2}%
}{b^{2}+b_{0}^{2}}\right)  ^{n}%
\end{equation}
where $j=l+3/2$. This overlap decreases, for $b>b_{0}$, for large values of
$n$ as
\begin{equation}
\left\langle \Phi_{nl}(b)\,|\,\Phi_{0l}(b_{0}\right\rangle \approx\left(
\frac{2b_{0}}{b}\right)  ^{j}\sqrt{\frac{n^{j-1}}{\Gamma\left(  j\right)  }%
}\exp\left(  -\left(  2n+j\right)  \frac{b_{0}^{2}}{b^{2}}\right)
\end{equation}
The larger is the ratio $b_{0}/b$, the faster the overlap tends to zero. When
the ratio is small, the overlap slowly approaches to zero. In the region of
$n$, where the overlap is negligibly small, the expansion coefficients
$C_{nl}^{N,b}\left(  k,b_{0}\right)  $ coincide with $C_{nl}^{N}\left(
k,b\right)  $, defined by the formula (\ref{eq:C022}).

The system of linear equations (\ref{eq:newregneumrecc}) can now be solved in
the following way. We start the three-term recurrence relation in a region of
$n$ where the overlap is negligibly small, and take the original asymptotic
$C_{nl}^{N}\left(  k,b\right)  $\ values as seeding values, then solve
(\ref{eq:newregneumrecc}) for $C_{nl}^{N,b}\left(  k,b_{0}\right)  $ by
stepping down the recurrence towards $n=0$.

Thus we are led to a formulation of the scattering problem where we represent
the solution as follows. The wave function has two essential components. In
the internal region the effects of the potential are felt. We assume it has a
finite range and that this component can be approximated by a finite
combination of the $L^{2}$ basis states. In the asymptotic region the solution
must match the boundary condition which we now express with the Bessel and new
regularized Neumann functions.%

\begin{align}
\Psi_{l}\left(  r|k\right)   &  =\Psi_{l}^{I}(r|k\,)\,+\Psi_{l}^{B}%
(r|k)-\tan\delta_{l}(k)\overline{\,\Psi}_{l}^{N}\left(  r\,|\,k,b_{0}\right)
\,\\
\Psi_{l}^{I}(r|k) &  \rightarrow0\,\,\,\,\,\,\text{for}\,\,\,r\rightarrow
\infty
\end{align}
In the algebraic representation this becomes%

\begin{align}
C_{nl}(k,b) &  =C_{nl}^{I}(k,b)+C_{nl}^{B}(k,b)-\tan\delta_{l}(k)\,\,C_{nl}%
^{N,b}(k,b_{0})\\
C_{nl}^{I}(k,b) &  \rightarrow0\,\,\,\,\,\,\text{for}\,\,\,n\rightarrow\infty
\end{align}
The equation for the internal wavefunction and the phase shift is obtained by
inserting (\ref{eq:C000}) into the Schr\"{o}dinger equation, projected onto
the basis set:%

\begin{equation}
\left\langle \,\,\Phi_{nl}(b)|\,T_{l}+V-E\,|\,\Psi_{l}^{I}(k)+\Psi_{l}%
^{B}(k)-\tan\!\delta_{l}\,\overline{\Psi}_{l}^{N}\left(  k,b_{0}\right)
\right\rangle \,\,=0\,
\end{equation}
Taking advantage of the properties of the asymptotic reference states
(\ref{eq:bessel}) and (\ref{eq:regNeumm}) this simplifies to:%

\begin{align}
&  \left\langle \Phi_{nl}(b)|T_{l}+V-E|\Psi_{l}^{I}(k)\right\rangle
\nonumber\\
&  -\tan\delta_{l}\left(  \beta_{0}\left\langle \Phi_{nl}(b)|\overline{\Psi
}_{l}^{N}\left(  k,b_{0}\right)  \right\rangle +\left\langle \Phi
_{nl}(b)|V|\overline{\Psi}_{l}^{N}\left(  k,b_{0}\right)  \right\rangle
\right) \nonumber\\
&  =-\left\langle \Phi_{nl}(b)|V|\Psi_{l}^{B}(k)\right\rangle
\end{align}
In the approach of \cite{kn:Heller1,kn:Yamani} the algebraic Schr\"{o}dinger
equation is approximated by assuming that the interaction matrix for potential
$V$ can be truncated at some large $n=N$. This $N$ defines a sharp boundary
between the internal and asymptotic regions in the coefficient space. The
resulting problem can be solved with $C_{nl}^{I}(k,b)=0$\ for $n\,\geq N$ .
One has an $N+1$ by $N+1$\ matrix equation in the unknowns$\{C_{0l}%
^{I},C_{1,l}^{I},\ldots,C_{N-1l}^{I},\tan\delta_{l}\}$:%

\begin{align}
&  \sum_{m=0}^{N-1}\left\langle \,\Phi_{nl}(b)|\,T_{l}+V-E\,|\,\Phi
_{ml}(b)\right\rangle \,C_{ml}^{I}(k,b)\nonumber\\
&  \;\;\;\;\;-\tan\!\delta_{l}\left(  \beta_{0}C_{nl}^{N,b}(k,b_{0}%
)+\sum_{m=0}^{N-1}\left\langle \Phi_{nl}(b)|V\,|\Phi_{ml}(b)\right\rangle
C_{ml}^{N,b}(k,b_{0})\right) \nonumber\\
&  =-\sum_{m=0}^{N-1}\left\langle \,\Phi_{nl}(b)|V\,|\,\Phi_{ml}%
(b)\right\rangle \,C_{ml}^{B}(k,b)
\end{align}%
\begin{figure}
[ptb]
\begin{center}
\includegraphics[
trim=0.000000in 0.000000in 0.211997in 0.423785in,
natheight=5.885900in,
natwidth=8.833200in,
height=7.8398cm,
width=12.334cm
]%
{PhasSQW_vs_b0.eps}%
\caption{$s$-wave phase shift for the square-well potential, obtained with
$b=3$ fm and varying renormalization widths $b_{0}$}%
\label{Fig:PhasSQW_vs_b0}%
\end{center}
\end{figure}
\begin{figure}
[ptbptb]
\begin{center}
\includegraphics[
trim=0.000000in 0.000000in 0.212880in 0.423196in,
natheight=5.885900in,
natwidth=8.833200in,
height=7.8419cm,
width=12.334cm
]%
{PhasYukawa_vs_b0.EPS}%
\caption{$s$-wave phase shift for the Yukawa potential, obtained with $b=3$ fm
and varying renormalization widths $b_{0}$}%
\label{Fig:PhasYukawa_vs_b0}%
\end{center}
\end{figure}
Convergence of the results is achieved by extending the interaction region
i.e. increasing $N$.

\section{Some examples}

In this section we present some detailed results for radial 1-dimensional
model potentials. We compare the application of the traditional JM
regularization procedure to the one introduced in section
\ref{sect:JMrevisited} for Gauss, exponential, Yukawa and square-well
potentials. We choose comparable\ parameters, in particular a depth of
$V_{0}=-80$ MeV and width $a=1.0$ fm, for all of these potentials.

All calculations are made in the standard JM approach, i.e. truncating the
potential matrix beyond the boundary condition matching point $N$
\cite{kn:Heller2,kn:Heller1,kn:Fil_Okhr}. For the traditional
regularization we consider the standard $C_{nl}^{B}(k,b)$ and $C_{nl}%
^{N}(k,b)$ asymptotic expansion forms, whereas for the new regularization
(J-Matrix Regularized (JMR) method) the $C_{nl}^{B}(k,b)$ and $C_{nl}%
^{N,b}\left(  k,b_{0}\right)  $ asymptotic forms are used. We have set the
matching point to $N=25$ in all cases. To obtain a fair comparison of the
convergence properties for the two regularization properties, we compare the
$s$-wave phase shifts to each other and to the \textquotedblleft
exact\textquotedblright\ result, obtained with the Variable Phase approach
\cite{kn:Calogero,kn:Babikov}.%
\begin{figure}
[ptb]
\begin{center}
\includegraphics[
trim=0.000000in 0.000000in 0.110931in 0.329568in,
natheight=9.552700in,
natwidth=6.933200in,
height=16.9382cm,
width=12.5471cm
]%
{PhaseGaussRegul.eps}%
\caption{Comparison of $s$-wave pahse shifts for JM and JMR for a Gauss
potential ($V_{0}=80$ MeV, $a=1.0$ fm). The matching point is $N=25$ in both
calculations. The Regularization parameter is $b_{0}=0.6$ fm.}%
\label{Fig:PhasesGauss1}%
\end{center}
\end{figure}
\begin{figure}
[ptbptb]
\begin{center}
\includegraphics[
trim=0.000000in 0.000000in 0.109545in 0.329407in,
natheight=9.493000in,
natwidth=6.933200in,
height=16.8305cm,
width=12.5471cm
]%
{PhaseExponRegul.eps}%
\caption{Comparison of $s$-wave pahse shifts for JM and JMR for an exponential
potential ($V_{0}=80$ MeV, $a=1.0$ fm). The matching point is $N=25$ in both
calculations. The Regularization parameter is $b_{0}=0.6$ fm.}%
\label{Fig:ExponPhases1}%
\end{center}
\end{figure}
\begin{figure}
[ptbptbptb]
\begin{center}
\includegraphics[
trim=0.000000in 0.000000in 0.109545in 0.329407in,
natheight=9.493000in,
natwidth=6.933200in,
height=16.8305cm,
width=12.5493cm
]%
{PhaseSQWRegul.eps}%
\caption{Comparison of $s$-wave pahse shifts for JM and JMR for a square-well
potential ($V_{0}=80$ MeV, $a=1.0$ fm). The matching point is $N=25$ in both
calculations. The Regularization parameter is $b_{0}=0.6$ fm.}%
\label{Fig:PhasesSQW1}%
\end{center}
\end{figure}
\begin{figure}
[ptbptbptbptb]
\begin{center}
\includegraphics[
trim=0.000000in 0.000000in 0.109545in 0.329407in,
natheight=9.493000in,
natwidth=6.933200in,
height=16.8305cm,
width=12.5493cm
]%
{PhaseYukawaRegul.eps}%
\caption{Comparison of $s$-wave pahse shifts for JM and JMR for a Yukawa
potential ($V_{0}=80$ MeV, $a=1.0$ fm). The matching point is $N=25$ in both
calculations. The Regularization parameter is $b_{0}=0.6$ fm.}%
\label{fig:PasesYukawa1}%
\end{center}
\end{figure}

In figures \ref{Fig:PhasSQW_vs_b0} and \ref{Fig:PhasYukawa_vs_b0} we show how
the solutions obtained with the new regularization (JMR) depend on its
defining value $b_{0}$. We have done this for both the Yukawa an square-well
potentials, as it is on these potentials that the effects are more noticeable.
It is also seen that the optimal value of $b_{0}$ strongly depends on the
functional form of the potential, more particularly on the behavior at the origin.

In figures \ref{Fig:PhasesGauss1}, \ref{Fig:ExponPhases1},
\ref{Fig:PhasesSQW1} and \ref{fig:PasesYukawa1} the comparison of the JM, JMR
and VPA $s$-wave phase shifts for all four model potentials is displayed, for
three different values of the oscillator length $b=1.0,\ 2.0$ and $3.0$ fm.
For uniformity the value for the regularization parameter was chosen to be
$b_{0}=0.6$ fm in all cases, as it is a near optimal value to limit the
differences between the renormalized $\overline{\Psi}_{L}^{N}\left(
r|k,b_{0}\right)  $ and true asymptotic Neumann $\Psi_{L}^{N}\left(
r|k\right)  $ functions in coordinate representation (see figure
\ref{Fig:PsiMinus1}).

We notice that, for all potentials, the results deviate more from the exact
(VPA) results with increasing $b$. This is an indication that a larger value
for the matching point $N$ should be chosen for truly convergent results. The
phase shifts obtained by JMR however remain much closer to the VPA results.
The potential effects are more pronounced for the square-well (figure
\ref{Fig:PhasesSQW1}) and Yukawa potential (figure \ref{fig:PasesYukawa1}).
The regularization effect is in all cases seen to be more important for large
values of $b$ where the convergence problem becomes an important issue.

It is also clear that convergence, both for JM and JMR, can be more easily
achieved for the Gauss and exponential potentials than for the square-well and
the Yukawa potentials. This undoubtedly has to do with the more regular
behavior at the origin of the former potentials. The JMR results can be
further improved if a more judicious choice is made for the regularization
parameter $b_{0}$. Indeed, from figures \ref{Fig:PhasSQW_vs_b0} and
\ref{Fig:PhasYukawa_vs_b0} it is seen that a value of $b_{0}$ around 0.4 fm
(square-well) or 0.2 fm (Yukawa) would improve the results even more drastically.

It should also be clear that combining the new regularization method
introduced here with other methods for improving convergence, e.g. using
semiclassical potential considerations such as \cite{2002PhRvL..88a0404V},
will carry the merits of both approaches, and can dramatically reduce the size
of potential matrices.

\section{Conclusions}

We have investigated the regularization procedure of the JM method. We have
shown that the differences between the regularized Neumann and Neumann
functions can have a significant impact on the convergence of the JM method.
We have proposed a new approach to the regularization. It is computationally
only slightly more involved but it yields significant improvement in the
convergence of the phase shifts obtained by the method. We have explicitly
demonstrated the latter point through calculations on several model potentials.

\appendix

\section{Overlap}

\label{Sect:Appendix}

To calculate the overlap between oscillator functions of different widths we
use the technique of generating functions. It is known that the function%
\begin{equation}
\Phi_{l}(r\,|\,\varepsilon,b)=\sqrt{\frac{2}{\Gamma\left(  l+3/2\right)  }%
}\frac{1}{b^{3/2}}\left(  1+\varepsilon\right)  ^{-(l+3/2)}\left(  \frac{r}%
{b}\right)  ^{l+3/2}\exp\left\{  -\frac{1}{2}\frac{1-\varepsilon
}{1+\varepsilon}\left(  \frac{r}{b}\right)  ^{2}\right\}  \label{eq:A101}%
\end{equation}
generates a complete set of the oscillator functions of width b%
\begin{equation}
\Phi_{l}(r\,|\,\varepsilon,b)=\sum_{n=0}^{\infty}\frac{\varepsilon^{n}%
}{\mathcal{N}_{nl}}\Phi_{nl}\left(  r,b\right)  \label{eq:A102}%
\end{equation}
where%
\begin{equation}
\mathcal{N}_{nl}=\sqrt{\frac{n!\Gamma\left(  j\right)  }{\Gamma\left(
n+j\right)  }}\,\;\;,\;\;j=l+3/2
\end{equation}
Using the relation%
\begin{equation}
\frac{\mathcal{N}_{nl}}{n!}\left[  \left(  \frac{d}{d\varepsilon}\right)
^{n}\Phi_{l}(r|\,\varepsilon,b)\right]  |_{\varepsilon=0}=\Phi_{nl}\left(
r|\,b\right)  \label{eq:A102a}%
\end{equation}
on matrix elements of the generator state, one can extract the oscillator
function matrix elements for definite quantum number $n$.

It is straightforward to calculate the overlap of two generating functions.
One finds%
\begin{equation}
\left\langle \Phi_{l}(\varepsilon,b)\,|\,\Phi_{l}(\widetilde{\varepsilon
},\widetilde{b})\right\rangle =\left[  \sqrt{\gamma}\frac{z}{\Delta}\right]
\,^{j}%
\end{equation}
where%
\begin{equation}
\gamma=\left(  \frac{\widetilde{b}}{b}\right)  ^{2},\qquad z=\frac{2}{\left(
1+\gamma\right)  },\quad\alpha=\frac{\left(  1-\gamma\right)  }{\left(
1+\gamma\right)  },\quad\Delta=1-\alpha\varepsilon+\alpha\widetilde
{\varepsilon}-\varepsilon\widetilde{\varepsilon} \label{eq:A104}%
\end{equation}
Closer inspection of the expression will reveal that it is actually symmetric
in the width parameters as it should be. By applying relation (\ref{eq:A102a})
to this overlap we obtain the required overlap of the oscillator functions:%
\begin{equation}
\frac{\emph{N}_{nl}}{n!}\frac{\mathcal{N}_{\widetilde{n}l}}{\widetilde{n}%
!}\left[  \left(  \frac{d}{d\varepsilon}\right)  ^{n}\left(  \frac
{d}{d\widetilde{\varepsilon}}\right)  ^{\widetilde{n}}\left\langle \Phi
_{l}(\varepsilon,b)\,|\,\Phi_{l}(\widetilde{\varepsilon},\widetilde
{b})\right\rangle \right]  _{\varepsilon=\widetilde{\varepsilon}%
=0}=\left\langle \Phi_{nl}(b)\,|\,\Phi_{\widetilde{n}l}(\widetilde
{b})\right\rangle
\end{equation}
It can be represented in the following form:%
\begin{align}
&  \left\langle \Phi_{nl}(|\,b)\,|\,\Phi_{\widetilde{n}l}(|\,\widetilde
{b})\right\rangle \nonumber\\
&  =\emph{N}_{nl}\mathcal{N}_{\widetilde{n}l}\left(  z\sqrt{\gamma}\right)
^{j}\alpha^{n+\widetilde{n}}\frac{\Gamma\left(  n+j\right)  \Gamma\left(
\widetilde{n}+j\right)  }{n!\widetilde{n}!\Gamma\left(  j\right)
\Gamma\left(  j\right)  }\nonumber\\
&  \sum_{k=0}^{\min\left(  n,\widetilde{n}\right)  }\frac{n!\widetilde
{n}!\Gamma\left(  j\right)  }{k!\left(  n-k\right)  !\left(  \widetilde
{n}-k\right)  !\Gamma\left(  k+j\right)  }\left[  -\frac{\left(  1-\alpha
^{2}\right)  }{\alpha^{2}}\right]  ^{k}\nonumber\\
&  =\emph{N}_{nl}\mathcal{N}_{\widetilde{n}l}\left(  z\sqrt{\gamma}\right)
^{j}\alpha^{n+\widetilde{n}}\frac{\Gamma\left(  n+j\right)  \Gamma\left(
\widetilde{n}+j\right)  }{n!\widetilde{n}!\Gamma\left(  j\right)
\Gamma\left(  j\right)  }\nonumber\\
&  \ _{2}F_{1}\left(  -n,-n\,;\,j\,;-\frac{\left(  1-\alpha^{2}\right)
}{\alpha^{2}}\right)
\end{align}
For the particular case where one of the quantum numbers is zero one finds:
\begin{align}
\left\langle \Phi_{nl}(|\,b)\,|\,\Phi_{\widetilde{n}l}(|\,\,\widetilde
{b})\right\rangle  &  =\mathcal{N}_{nl}\left(  z\sqrt{\gamma}\right)
^{j}\alpha^{n+\widetilde{n}}\frac{\Gamma\left(  n+j\right)  }{n!\Gamma\left(
j\right)  }\nonumber\\
&  =\left(  \frac{2b\widetilde{b}}{b^{2}+\widetilde{b}^{2}}\right)  ^{j}%
\sqrt{\frac{\Gamma\left(  n+j\right)  }{n!\Gamma\left(  j\right)  }}\left(
\frac{b^{2}-\widetilde{b}^{2}}{b^{2}+\widetilde{b}^{2}}\right)  ^{n}%
\end{align}


\end{document}